\documentclass[epj]{svjour}
\usepackage{graphicx,amsmath,amssymb}

\def\p{\partial}
\def\be{\begin{equation}}
\def\ee{\end{equation}}
\def\bp{\bgroup \bf p\egroup}
\def\bn{\textbf{n}}
\def\bv{\textbf{v}}
\def\bi{\bgroup \bf e\egroup}

\def\Tr{\mathop{\rm Tr}\nolimits}
\let\phi\varphi
\let\div\odiv
\def\div{\mathop{\rm div }\nolimits}

\def\({\biggl(}
\def\){\biggr)}
\def\[{\biggl[}
\def\]{\biggr]}

\def\mynote#1{\bgroup\hskip1ex$\big($\bf #1$\big)$\egroup}
\def\mymargin#1{\bgroup\rule{1pt}{8pt}\marginpar{#1}}

\begin{document}

\title{Cell motility: a viscous fingering analysis of active gels}
\author{M. Ben Amar\inst{1} \and O. V. Manyuhina\inst{1} \and G. Napoli\inst{2}%
}                     
%
%
\institute{\inst{1}{Laboratoire de Physique Statistique, Ecole Normale Sup\'erieure, UPMC  
Univ Paris 06, Universit\'e Paris Diderot, CNRS, 24~rue Lhomond, 75005  
Paris,  France}\\ \inst{2}{Dipartimento di Ingegneria dell'Innovazione Universit\`a del Salento, Via per Monteroni -- Edificio ``Corpo O'', I-73100 Lecce, Italy}}
\date{Received: date / Revised version: date}
%
\abstract{The symmetry breaking of the actin network from radial to longitudinal symmetry  has been identified as the major mechanism for keratocytes (fish cells) motility on solid substrate.  For strong friction coefficient, the two dimensional  actin flow  which includes the polymerisation at the edge and depolymerisation in the bulk can be modelled as a Darcy flow,  the cell shape and dynamics  being  then modelled by standard complex analysis methods. We use the theory of active gels to describe the orientational order of the filaments which varies from the border to the bulk.  We show analytically that the reorganisation of the cortex is enough to explain the motility of the cell  and find the velocity as a function of the orientation order parameter in the bulk.}

\PACS{
      {87.17.Jj}{Cell locomotion; chemotaxis} \and
      {87.17.Rt}{Cell adhesion and cell mechanics}\and 
      {47.20.Gv}{Viscous and viscoelastic instabilities}\and
      {47.20.Ky}{Nonlinearity, bifurcation, and symmetry breaking}\and
      {47.20.Ma}{Interfacial instabilities}  
     } 
%

\maketitle

\section{Introduction}\label{sect:intro}

\begin{figure}
\centering
\raisebox{37mm}{A)}\kern10pt\includegraphics[height=40mm]{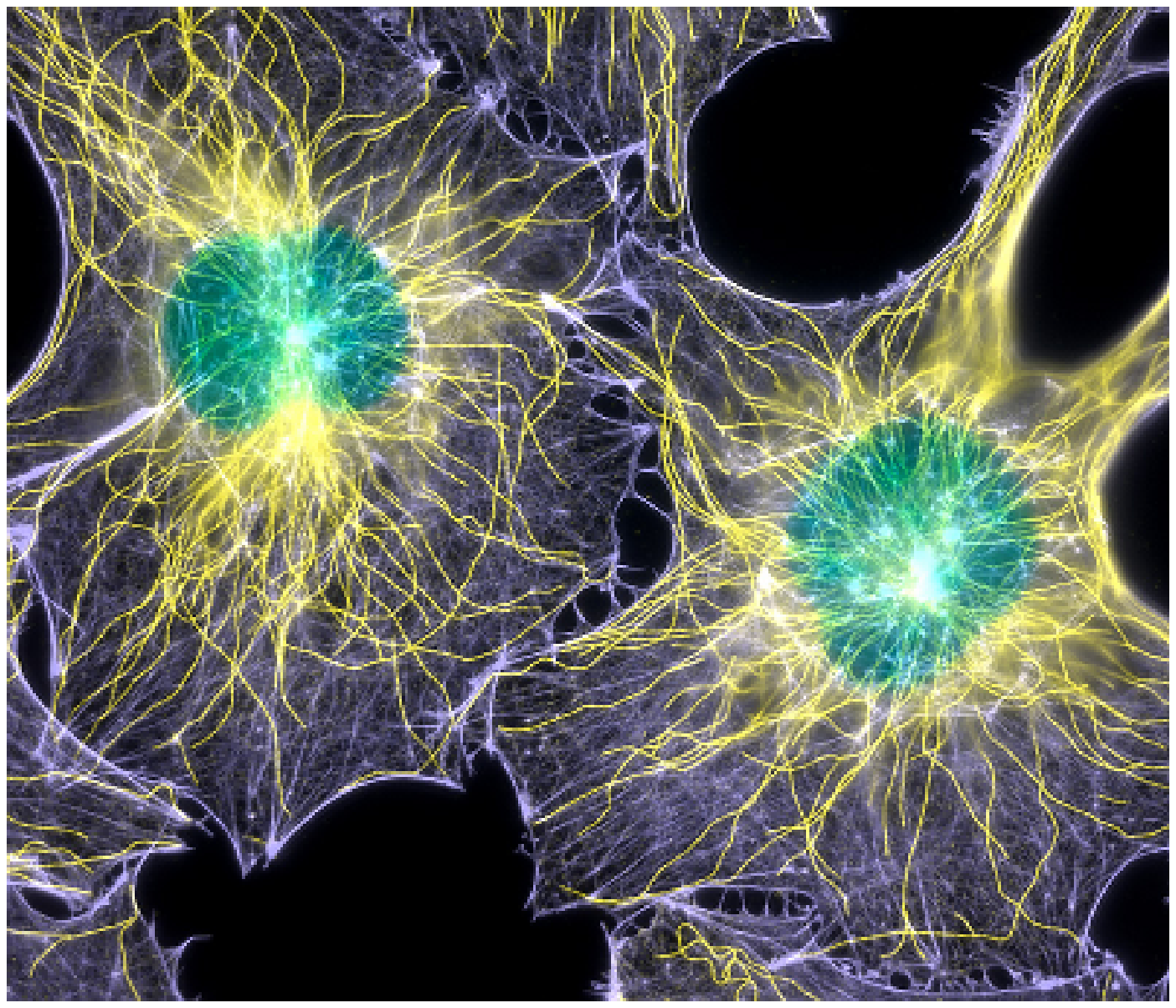}\hfil
\raisebox{37mm}{B)}\kern10pt\includegraphics[height=40mm]{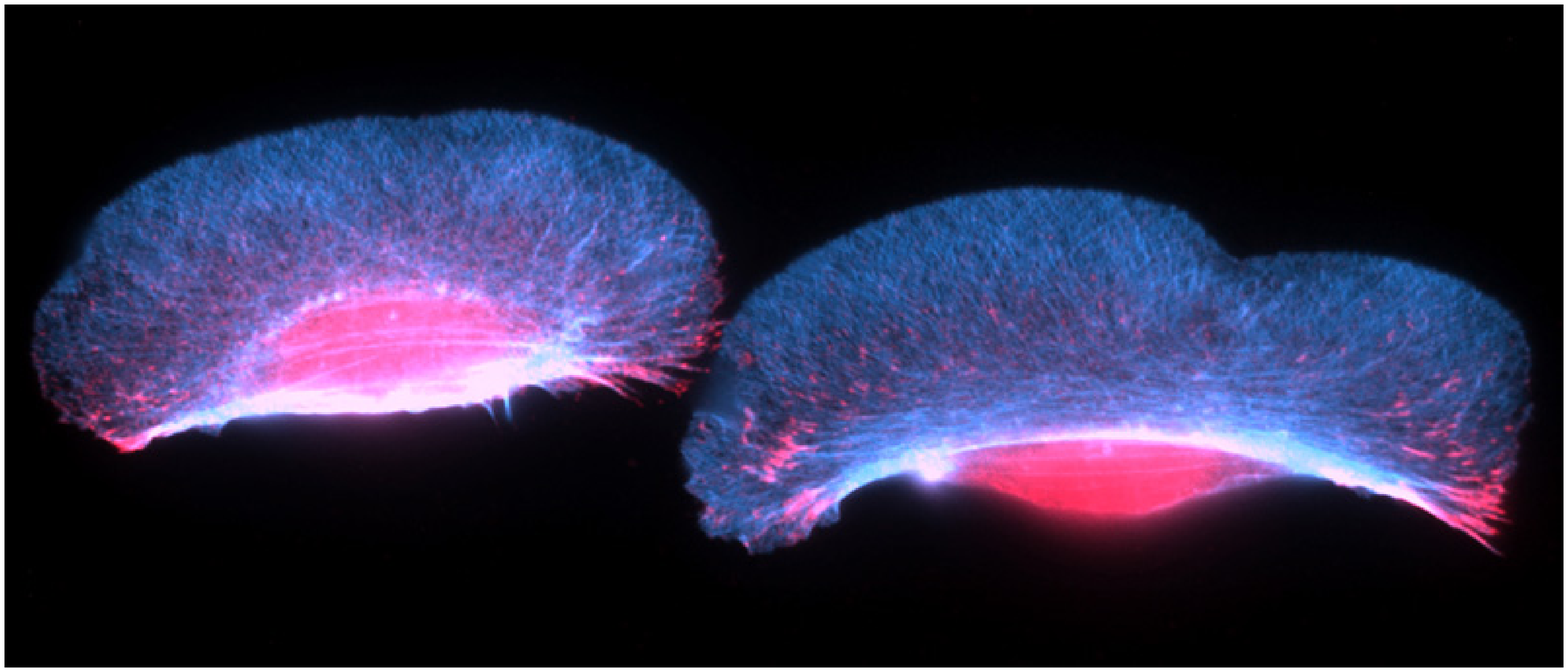}
\caption{A) Mouse fibroblasts stained for actin (blue), microtubules (yellow), and nuclei (green). B) Migrating goldfish skin keratocytes stained for actin (blue) and adhesion sites (red). Courtesy Professor Torsten Wittmann, University of California (http://www.ucsf.edu/science-cafe/conversations/wittmann/).}
\label{fig:keratocyte}
\end{figure}

Three polymers, actin filaments, microtubules and intermediate filaments, are responsible of the eukaryotic cell integrity~\cite{fletcher}. Together with molecular motors, these filaments accomplish tasks necessary to the cell survival and cell mitosis. Microtubules are known to transport intracellular molecules, chromosomes and organelles inside the cells. They play also an important role in multi-cellular organisms  during embryogenesis. As an example, for  C-elegans  at some stage of the development, they modify the shape of the embryo from an ovoid to a worm-like shape by redistributing the stress originating from the contraction of the actin--myosin network inside  the epithelial cells~\cite{ciarletta}. Assembly and disassembly of these filaments~\cite{pollard} (microtubule and actin filaments, see  fig.~\ref{fig:keratocyte}A) produce forces, generate tension and are responsible of the cell shape deformations.

Here we focus on the crawling of cells like keratocytes (fish epithelial cell) or fibroblasts. By occurring on a solid substrate, in vitro, this process induces the protrusion of a lamellipodium mainly due to the polymerisation of the actin network in the direction of the displacement. Observations in vitro reveal four well identified stages: the formation of this lamellipodium, the adhesion to the substratum, the retraction of the rear, then the de-adhesion. They also clearly indicate a shape bifurcation of the whole flat cellular shape from a static position where the cell is mostly round to a crescent shape which accompanies the displacement. Visualization of the actin cortex via fluorescent speckle microscopy shows a reorganization of the actin network inside the cell in concordance with this shape symmetry breaking.  For keratocytes, F-actin moves from the leading edge to the cell core at a rate about 25-60 nm/s. The flux is known to be faster at the periphery, decreasing in the  center. In motile keratocytes (see fig.~\ref{fig:keratocyte}B), F-actin moves rewards slowly as the cell moves rapidly forwards. In this case the F-actin flow decreases to 10 nm/s. The biological processes involved in cell motility are complex and many actors play a decisive role in this migration process like molecular motors  (mostly myosin) responsible of the deformation of the actin cortex or the Wiskott-Aldrich proteins (WASP) which regulate the polymerization process~\cite{alberts} at the level of the lipid membrane.  

Our goal is not to enter into all the details of the microscopic biological mechanisms but to explain the existence of a displacement when the cortex of actin is reorganized by the molecular motors.  We adopt a macroscopic viewpoint by treating the F-actin network as a nematic two-dimensional gel put in a situation out of equilibrium by ATP hydrolysis. We will extend the approach of Callan {\it et al.}~\cite{andrew} limited to  the  circular description  of the actin network so to static cell shapes by taking into account the evolution of the orientation of the filaments as the cell moves. This requires a modification of the constitutive equations of  active gels, including nonlinearities in the free-energy to explain the localization of the cortex at the cell border.  The tensorial resulting balance equations are coupled to boundary conditions resulting from mechanical equilibrium and observation of the actin network. As an example, it is known that the filaments approach the lipid membrane quite tangentially (with some small angle of order twenty degrees~\cite{fletcher,andrew}). The mathematical solutions of these equations with these boundary conditions lead to a free-boundary problem where the cell shape cannot be fixed a priori. Treating the disorientated cortex at the cell border as a boundary layer in both cases (static and motile) with an isotropic core (static case) or a fully oriented core (motile case), we succeed to formulate mathematically the shape equation with the Schwarz function~\cite{cummings}, a technique used for shape drops in hydrodynamics and wetting~\cite{benamar1}. The actin cortex, once treated as a boundary layer, induces  a modification of the boundary conditions for  the hydrodynamic flow.  In particular we explain the destabilization of the static cell for an increase of tensile activity of the molecular motors and the motility of the cell, after reorganization of the core for cells which remain quasi-circular.
 
The paper is organized as follows. First we show that the process of polymerization-depolymerisation induces a flow which can be approximated by some Darcy flow when  the friction on the substrate is important and we establish the boundary conditions of Neumann and Dirichlet type at the origin of the free-boundary problem. In section 3 we present the complex analysis formulation~\cite{cummings} and prove mathematically that the Callan et al~\cite{andrew} approach cannot explain the cell motility. Section 4 is a reminder of the theory of active gels~\cite{kruse:2005,julicher:2007,salbreaux:these} with application to the polar geometry (2D cylindrical geometry).  Section 5 considers the case where the actin cortex is limited to a boundary layer at the cell periphery with a melted and oriented core. Section 6 establishes the modification of the boundary conditions while in  section 7 we show that at leading order our theory is enough to explain motility for a circular cell with oriented core.

\section{Formulation of the problem}\label{sect:problem}

\subsection{The flow origin}

We assume  that  the shape of the drop is fixed by the actin flow in interaction with the lipid membrane~\cite{andrew}. It is the result of the depolymerisation of the actin filaments which is isotropically distributed  in the bulk (called $\Omega$ in the following) and of the polymerisation which occurs at the boundary ($\partial \Omega$). Such an assumption is inspired by the experimental observations~\cite{yam}: a flow of actin is visualized and reveals that it is directed  towards some center in the cell. Such assumption is consistent with the fact that circular cells of constant thickness do not move on the substrate. Indeed, we assume that the actin filaments arrive at the cell border with some orientation with the normal and touch it with the pole plus. So polymerisation occurs at the border which becomes a sink  of actin monomers  contrary to the bulk which is a source. The conservation of mass is represented by the continuity equation:
\begin{equation}
\label{balance}
\frac{\partial \rho}{\partial t}+ \nabla\cdot(\rho {\bf v})=-\rho k_d+  \rho {\cal F}  V_p\vec {\bf n}\, \delta(  ({\bf r}- {\bf r}_{int}) \cdot {\bf n}),
\end{equation}
where $\rho$ is the actin filament concentration,  moving with a local  velocity ${\bf v}$ .  The actin gel is assumed incompressible and there is a conservation of the total number of actin monomers (either free or organised in filaments or bundles). $\rho$ is assumed constant inside the  cell. Indeed, since our model is two-dimensional  and since there exists some thickness variation especially at the border~\cite{yam}, this hypothesis may be questionable but we will discard this thickness effect and keep a purely two-dimensional model.  We call  $V_p$  the polymerisation velocity, $ {\bf r}_{int}$  the position of a  point on the interface and $\bn$ the normal at this point to the interface.  A dimensionless shape factor ${\cal F}$ indicates that the density of actin filaments at the border may vary from point to point. It is typically a function of the local curvature or its even derivatives. We assume that  this shape factor ${\cal F}$ is normalized to $1$ for the circle which has a constant curvature. We can tune  ${\cal F}$  if we have biological informations. For example  Yam {\it et al}~\cite{yam} mention that when cells become motile, they are no longer circular and the concentration of  actin touching the border is higher at the front than at the rear.  By taking into account a small volume element centered along the cell border and integrating this equation, we derive 
\begin{equation}\label{bound}
 v_n= {\bf v}\cdot{\bf n} = U {\bf e}_{y} \cdot {\bf n}-{ \cal F}  V_p,
\end{equation}
 In the bulk ($\Omega$), depolymerisation of actin filaments gives  
\begin{equation}\label{div}
\nabla\cdot  {\bf v}=-k_d.
\end{equation}
Moreover, mass conservation of actin filaments imposes that:
\begin{equation}
k_d \int_{\Omega} dx\,dy=V_p \int_{\delta \Omega}\!\! { \cal F}\, ds,
\end{equation}
where $s$ is a curvilinear coordinate along the contour $\delta \Omega$. As an example, for a circular cell of radius $R$, we get  $k_d R=V_p/2$, which coincides with the result derived in~\cite{andrew}. Note that this mass conservation has to be imposed at each step of the cell evolution. It is a constraint which must be satisfied by the shape solution. For non-steady dynamics of the motility process, for transitory states or stability studies for example, eq.~(\ref{bound}) is easily modified by introducing the local velocity of the boundary ${\bf V}_{\delta \Omega}$ so we get 
\begin{equation}\label{bound1}
   {\bf v} \cdot{\bf n}  +{ \cal F}V_p= {\bf V}_{\delta \Omega}\cdot {\bf n}.
\end{equation}

\subsection{The flow equation}

We neglect here inertia and  the Stokes equation inside the cell is given by
\begin{equation}\label{stokes}
- \nabla P -\xi  {\bf v}+\eta  {\rm \Delta}  {\bf v} =0,
\end{equation}
if  the cell is assumed to be isotropic and homogeneous. Indeed, the cell contains filaments with molecular motors able to transmit stresses which can modify  eq.~(\ref{stokes}). When these filaments are completely disoriented, we will show in section~\ref{sect:nematic}, that eq.~(\ref{stokes}) is not modified although the system is active. It means that inside the cell, there exists permanent stresses induced chemically by ATP hydrolysis.   When  anisotropy of the actin network will be introduced, eq.~(\ref{stokes}) will be  modified (see section 6).  In the limit when the friction coefficient $\xi$ is strong, we can neglect the viscosity $\eta$ and recover from eq.~(\ref{stokes}) the  Darcy law for the flow inside the drop. First we consider the full problem, then the limiting cases. 

The flow velocity during motility initiation has been visualised  experimentally using multiframe correlation tracking. Initially radial and centripetal, it  becomes oriented along the axis of displacement. To satisfy  eq. (\ref{div}), we take into account these two cases. We define the stream function $\Psi$ which allows to derive the incompressible part of the velocity field.  In a Cartesian coordinate system, we have 
\begin{equation}\label{eq:stream}
v_x=-\frac{ k_d}{2} x+ \frac{\partial \Psi}{\partial y}, \qquad v_y=-\frac{ k_d}{2} y-\frac{\partial \Psi}{\partial x},
\end{equation}
for a flow field mostly radial, and for a flow field mainly oriented along the $y$ direction 
\begin{equation}\label{eq:stream1}
v_x=\frac{\partial \Psi}{\partial y}, \qquad v_y=-\frac{\partial \Psi}{\partial x}- k_d y +V_0.
\end{equation}
being $V_0$ a constant divergence-less velocity. 
The Stokes equation eq.~(\ref{stokes}) is then transformed into: 
\begin{align}\label{stokes1}
-\frac{\partial P}{\partial x}+\eta {\rm \Delta} \frac{\partial \Psi}{\partial y}-\xi \frac{\partial \Psi}{\partial y}&=0 \\
\label{stokes2}
-\frac{\partial P}{\partial y}-\eta {\rm \Delta} \frac{\partial \Psi}{\partial x}+\xi \frac{\partial \Psi}{\partial x}&=-\xi k_d y+\xi V_0.
\end{align}
Once equation~(\ref{stokes1}) is differentiated with respect to $y$ and eq.~(\ref{stokes2}) with respect to $x$, the pressure $P$ can be eliminated and we get 
\begin{equation}\label{complete}
\eta {\rm \Delta}^2 \Psi-\xi {\rm\Delta} \Psi=0.
\end{equation}
When the  active stress $\sigma^{act}_{ij}=-\zeta \delta \mu \delta_{ij}$  is introduced  isotropically,  it modifies   the boundary conditions. The material coefficient  $\zeta$  characterises the activity of the actin--myosin network inside the cell, being negative  for contractile  motors and positive for tensile motors;  $\delta \mu$ is the difference of the chemical potential between Adenosine triphosphate (ATP) and hydrolysis products. 
Let us now fix the boundary conditions: i) the normal component of the velocity at the boundary (eq.~(\ref{bound})) must be equal to the translational velocity of cell along the  normal, ii) the other conditions concern the mechanical balance of forces on the interface $\partial \Omega $ both  in the normal and the tangential directions
\begin{gather}\label{normalstress}
\sigma_{nn}= -P +2 \eta \frac{\partial v_n}{\partial n}=\zeta \delta \mu,\\
\sigma_{ns}= \eta \( \frac{\partial v_n}{\partial s}+\frac{\partial v_s}{\partial n}-\vert \kappa\vert v_s\)=\zeta \delta \mu.\label{eq:tangentstress}
\end{gather}
$v_n$, $v_s$ being the velocity normal and parallel to the boundary  $\delta \Omega$ while  
$\kappa$ is the local curvature. If we introduce a tension or capillarity effect, we must modify  eq.~(\ref{normalstress}) and add a contribution proportional to the curvature of the interface (Laplace law for the pressure jump). In case of the Laplacian flow with $\eta=0$, eq.~(\ref{eq:tangentstress}) should be ignored. 

For a simple cell geometry as the circular one, the shape equation can be derived by assuming the radial geometry for $\Psi$ as done in~\cite{andrew}. Nevertheless, sometimes this geometry, which is assumed {\it a priori}, cannot allow to satisfy all the equations because we face a free-boundary problem, which is especially difficult to analyse for Darcy and Stokes flows~\cite{cummings}.  Free boundary problems can be solved either by complex analysis~\cite{cummings}  or with the help of the Green function techniques. The latter case requires numerics while complex analysis is more elegant. Unfortunately, none of these two techniques is sufficient to give a solution of the complete problem in the case of eq.~(\ref{complete}) which couples a Laplacian and a Bi-Laplacian.  One possibility may be that the friction coefficient is large, so that the viscosity effect is located only at the boundary layer around the border of our cell. In this case, it will be possible to prove that effects of this viscous boundary layer result only in a modification of the boundary conditions by introducing ``a capillary term''~\cite{benamar}. 

Let us consider first the case without viscosity: the dimensionless parameter $\tilde \eta= \eta/ (\xi R^2)$~\cite{andrew} ($R$ being a radius of a circular cell at rest) goes to zero and we can define a velocity potential $\Phi^{(0)}$ equal to $-P/\xi$ . This velocity field satisfies the Poisson equation because of the depolymerisation effect. So we have ${\rm \Delta} \Phi^{(0)}=-k_d$ (equivalent to eq.~(\ref{div})) and two boundary conditions for the normal velocity (eq.~(\ref{bound})) and the other for the normal force  balance (eq.~(\ref{normalstress})) with $\eta=0$
\begin{gather}\label{vitnorm}
v_n = {\bf v} \cdot {\bf n}=\frac{\partial  \Phi^{(0)}}{\partial  {\bf n}}=U \cos\vartheta-V_p,\\
P=-\zeta \delta \mu + \mbox{ capillary terms}.
\end{gather}
The capillary terms include {\it e.g.} the tension, which exists in the lipid membrane and that the membrane exerts on the cell. Note that this problem has some mathematical similarity with the treatment of the sliding drop in the situation of partial wetting  given by Ben Amar {\it et al.}~\cite{benamar1} studied with the Schwarz function technique that we present now.

\section{Complex analysis formulation}

Here we develop the complex analysis formulation and Schwarz function technique which is well adapted to bounded domains. Assuming the strong  friction limit,  the Darcy law provides the velocity potential $\Phi^{(0)}$.  Let us write the boundary conditions using complex notations, assuming a  pressure jump at the boundary due to capillarity,
\begin{align}\label{neu}
\frac{\partial  \Phi^{(0)}}{\partial s}&=  {\bf t} \cdot \nabla \Phi^{(0)}=\frac{\gamma}{\xi}\frac{\partial^2 \vartheta}{\partial s^2},\\
\label{dir}
\frac{\partial \Phi^{(0)}}{\partial {\bf n}} &= {\bf n} \cdot \nabla  \Phi^{(0)}= (U \cos \vartheta-V_p)+\partial_t \Omega \cdot {\bf n},
\end{align}
where $\vartheta$ is the angle between the $y$-axis and the normal ${\bf n}$ to the interface or the angle between the $x$-axis and the tangent ${\bf t}$  to the interface.  The last term in eq.~(\ref{dir}) takes into account the fact that the border can fluctuate. By combining these two boundary conditions, we get 
\begin{equation}
2 \frac{\partial  \Phi^{(0)}}{\partial z}= \frac{\partial  \Phi^{(0)}}{\partial x}-i \frac{\partial  \Phi^{(0)}}{\partial y} = \( \frac{\partial  \Phi^{(0)}}{\partial s}-i \frac{\partial  \Phi^{(0)}}{\partial {\bn} }\) e^{- i \vartheta}=i V_p e^{-i \vartheta} -i\frac{U}{2} (1+  e^{- 2 i \vartheta}) + i \frac{\gamma}{\xi} \frac{\partial^2 \vartheta}{\partial s^2} e^{-i \vartheta}.
\end{equation}

We define the Schwarz function of the contour~\cite{benamar1} such that $\bar z\equiv g(z)$, where $z=x+iy$ is the complex variable and $\bar z=x-iy$ denotes the complex conjugate. By using the relation
\begin{equation}
 \frac{d \bar z}{ds}=g'(z) \frac{d z}{ds}= e^{-i \vartheta},
\end{equation}
we find that $e^{-i \vartheta}= \sqrt {g'(z)}$. So finally the complex derivative of the potential  along the interface is 
\begin{equation}
\label{main}
2\frac{\partial {  \Phi^{(0)}}}{\partial z}=i V_p\sqrt {g'(z)}+i\frac{\gamma}{2\xi} \frac{d}{dz} \(\frac{g''}{g'^{3/2}}\) - i \frac{U}{2}(1+g'(z)) +\frac{1}{2} \partial_t g.
\end{equation} 
Since ${\rm \Delta} \Phi^{(0)}=-k_d$ the velocity potential can be written as
\begin{equation}\label{eq:Phi0}
\Phi^{(0)}(x,y)=-\frac {k_d}4 z \bar z+ \frac 1 4 (H(z)+\bar H(\bar z)),
\end{equation}
where $H(z)$ is an arbitrary unknown function of the complex variable $z$. Combining these two results we find the interface equation that we extend to the whole volume of the drop
\begin{equation}
\label{schwarz}
-k_d g(z) + H'(z)=2 i V_p\sqrt {g'(z,t)}+i \frac{\gamma}{\xi} \frac{d}{dz} \(\frac{g''}{g'^{3/2}}\) -i U(1+g'(z)) +\partial_t  g(z).
\end{equation}
Let us consider the case now where the flow field is along the $y$ axis. As suggested by the experimental results of  \cite{yam},
we decompose the flow into a constant velocity $V_0$ and a flow with non-zero divergence contribution according to eq.(\ref{div}).   Then the flow potential becomes
\begin{equation}\label{eq:Phi1}
\Phi^{(0)}(x,y)=-\frac{k_d}{2} y^2+V_0 y+ \frac{1}{4} (\tilde H(z)+\bar{\tilde H}(\bar z))=\frac{k_d}{8}(z-\bar z)^2+ \frac{V_0}{2i} (z-\bar z) +  \frac{1}{4} (\tilde H(z)+\bar{ \tilde H}(\bar z)),
\end{equation}
which does not modify eq.~(\ref{schwarz}).

To simplify the notations, we have mentioned only the spatial dependance of the functions $\Phi^{(0)}, g(z)$  and its derivatives of first order $g'(z)$ and second order $g''(z)$ but it is obvious that in a non-steady problem they are time dependent as well. This equation has been established at the boundary $\partial \Omega$. We extend it to the whole domain $ \Omega$ by following the standard strategy for free-boundary problems. Since $\Phi^{(0)}$ represents a physical quantity, it cannot have singularities inside the domain $\Omega$ but the Schwarz function and its derivative  have  singularities inside $ \Omega$.  So we need to solve this equation with an arbitrary regular analytical function inside the drop $H(z)$ imposing  that  $g(z)$ is a Schwarz function that is: $z=\bar g(g(z))$ for arbitrary $z$ inside the drop. We can show straightforwardly that the static disc with $U=0$  is a solution of the eq.~(\ref{schwarz}). Since the Schwarz function of a circle of radius $R$ is $g(z)=R^2/z$, $g'(z)=-R^2/z^2$ so $\sqrt{g'(z)}=iR/z$, we recover $k_d R=2 V_p$.   Capillarity gives no contribution in this case. Moreover, it is absolutely not obvious to find a traveling wave-solution. Indeed, the leading order singularity related to the velocity $U$ is of the order of $g'$, which has no counterpart in eq.~(\ref{schwarz}), and thus cannot be cancelled~\cite{benamar1}. It means that such a circular drop cannot move with a constant velocity $U$. Our aim is then to analyse more deeply the polymerisation-depolymerisation process of the actin cortex localized below the lipid membrane  in order to modify the model which is too much simplistic. Note that if the polymerisation process depends explicitly on the polar angle, then the circular moving drop (with velocity $U$ along the $y$-axis)  is a solution of our problem. Indeed, by transforming $V_p$, which is the rate of polymerisation, into $V_p(1+ a \cos \vartheta)$ we get the following  flow equation:
\begin{equation}\label{schwarz1}
-k_d g(z) + H'(z)=2 i V_p \sqrt {g'(z,t)} + i V_p a (1+ g'(z)) +i \frac{\gamma}{\xi} \frac{d}{dz} \(\frac{g''}{g'^{3/2}}\) -i U(1+g'(z)) +\partial_t  g(z),
\end{equation}
and we can identify the drop velocity $U$  as the polymerisation rate times the anisotropy coefficient. Nevertheless, this cannot explain the biological processes at the origin of the bifurcation in shape motility. In the next section we  analyse the stability of such solutions. 
  
\subsection{Linear stability of the static circular solution}

The loss of stability of the static solution may explain the shape bifurcation and the cell motility. The Schwarz function technique is very efficient for performing such analysis avoiding complicated geometric calculations. It is why we present first the method performed  in~\cite{andrew}, then the method derived from complex analysis.

\subsubsection{Classical method}

The border is assumed to fluctuate around the circle and its equation is given by
\begin{equation}
r=R \big(1+\epsilon_n e^{\Omega_n t} \cos n\vartheta\big).
\end{equation}
The velocity potential $\Phi^{(0)}$ up to the leading order is
\begin{equation}
 \Phi^{(0)}(r,\vartheta)=-\frac{k_d}{4 } r^2+\frac{1}{4}  a_n e^{\Omega_n t}  r^n \cos n\vartheta.
\end{equation}
The Laplace equation at the boundary $\partial \Omega$  becomes 
\begin{equation}
 \Phi^{(0)}(r,\vartheta)=-\frac{k_d}{4} R^2 -\frac{k_d R ^2}{2} \epsilon_n e^{\Omega_n t} \cos{n\vartheta}+\frac{1}{4} a_n e^{\Omega_n t}  R^n \cos n\vartheta=-\frac{\gamma}{\xi} \frac{1}{R} (n^2-1) \epsilon_n e^{\Omega_n t}\cos n\vartheta,
\end{equation}
which allows to find the relation between $a_n$ and $\epsilon_n$, given by
\begin{equation}
a_n=\frac{4 \epsilon_n}{R^n}\( \frac{k_d}{2}R^2-\frac{\gamma}{\xi R} (n^2-1)\).
\end{equation}
Up to the linear order, the normal remains along the radius, so the linear expansion of eq.~(\ref{dir}) becomes
\begin{equation}
-\frac{k_d R}{2} \epsilon_n e^{\Omega_n t} \cos n\vartheta+\frac{n}{4}  a_n e^{\Omega_n t}  R^{n -1}\cos n\vartheta=\Omega_n R \epsilon_n e^{\Omega_n t} \cos n\vartheta,
\end{equation}
and finally the dispersion relation is
\begin{equation}\label{growthlinear}
\Omega_n=(n-1)\frac{V_p}{R}\(1-\frac{\gamma}{V_p \xi R^2} n(n+1)\).
\end{equation}

\subsubsection{Stability analysis with the Schwarz function method}

In the presence of small distortions at the contour $\partial \Omega$, the Schwarz function is given up to linear order by 
\begin{equation}
g(z)=\frac{R^2}{z}\big(1+\epsilon_n e^{\Omega_n t} (z^n +z^{-n})\big),
\end{equation}
and $H'(z)=b_n (n-1) e^{\Omega_n t} z^{n-1}$. The elimination of the singular mode $z^{-n}$  in eq.~(\ref{schwarz}) gives the following dispersion relation 
\begin{equation}
-k_d R^2\epsilon_n z^{-n-1} =-V_p R(n+1)\epsilon_n z^{-n-1}+\frac{\gamma}{\xi R} n(n^2-1) z^{-n-1}+\Omega_n R^2\epsilon_n z^{-n-1},
\end{equation}
which is exactly the same as in eq.~(\ref{growthlinear}).
Such analysis shows that the process of polymerisation-depolymerisation of actin destabilizes the cell when it is static but the tension in the lipid membrane at the origin of the capillary effect may be enough to stabilize the cell for small radius and moderate friction represented by $\xi$. Clearly the lipid membrane may readjust its tension: it has been experimentally shown that  the actin cortex strongly interacts with the lipid membrane~\cite{fletcher}.

In summary, this method gives an easy way to prove that a prescribed 2D-shape is a possible solution or cannot be a solution to a free-boundary problem. Predictions of more complex shapes than the circle is much more unlikely due to the difficulty to satisfy the requirement $z= \bar g(g(z))$.  The algebra for linear stability analysis close to an exact solution is straightforward with this method, as shown above. Anyway, we cannot have a motile cell with these boundary conditions. Our purpose  now is to take into account the anisotropy of the actin network to modify these boundary conditions.

\section{Modelling of active gels as two-dimensional ordered fluids}\label{sect:nematic}

\subsection{How to introduce the anisotropic nematic tensor}

We describe the actin filaments as nematic objects with  a local averaged orientation represented by  a unit vector $\bp$ called the director varying  from point to point inside the  cell.  The order inside  the cell is measured by the  orientational order parameter $q$. When   $q=0$ the filaments are randomly oriented and when $q=1$ they are perfectly ordered. Therefore, similar to nematic liquid crystals,  actin filaments may be described by a tensorial order parameter~\cite{pieranski,degennes:book}
\be\label{eq:Qtensor}
{\bf Q} = q \bigg(\bp\otimes \bp-\frac{\bf I}2\bigg) \quad\mbox{or}\quad Q_{ij} = q \big(p_ip_j-\delta_{ij}/2\big), 
\ee
where $\bf I$ is the two-dimensional identity tensor. One should remember that ${\bf Q}$  is a two-dimensional traceless symmetric tensor. Moreover, there is an intrinsic symmetry of the problem, namely the tensorial order parameter ${\bf Q}$, associated with negative $q$ and director ${\bf p}$, coincides with the one associated with positive $q$ and director orthogonal to ${\bf p}$. Thus, without loss of generality, we restrict our analysis to non-negative values of $q$.

The hydrodynamic equations for the order parameter $Q_{ij}$ of actin filaments and velocity of the gel can be regarded as an extension of the continuum theory for nematic liquid crystals with tensorial order~\cite{olmsted,sonnet} towards the theory of active gels~\cite{kruse:2005,julicher:2007,salbreaux:these}. In the cell cytoskeleton, the  molecular motors (mostly myosin II)  move along the actin filaments and   modify the organization of the actin network~\cite{fletcher} allowing the deformation of the cell and its possible displacement on a substrate. These  active motors  transform the chemical energy coming from Adenosine triphosphate (ATP) hydrolys into mechanical energy generating a nonzero active stress $\sigma^{act}_{ij}=-\zeta\delta\mu (Q_{ij}+\delta_{ij})$. According to  the sign of $\zeta$, negative (positive) the stress acting on the actin network is contractile (tensile). The coupled equations of motion for the order parameter  ${\bf Q}$ and the velocity field ${\bf v}$ are given in a macroscopic approach by~\cite{olmsted,kruse:2005,julicher:2007,salbreaux:these}  
\begin{align}
\xi v_i=\sigma_{ij,j},\label{eq:sigma}\\
\frac {D Q_{ij}}{Dt} = Q_{ij,k} v_k &= \beta_1 u_{ij} +\frac 1 {\beta_2} H_{ij} -\lambda\delta\mu Q_{ij},\label{eq:Q}
\end{align}
where $\sigma_{ij}$ represents the stress tensor, $v_{i,j}$ the velocity gradient and $u_{ij}$ the deviatoric part of the velocity gradient: $u_{ij}=(v_{i,j}+v_{j,i})/2 -  (v_{k,k}/2) \delta_{ij}$. The operator $D/Dt$ denotes the convective time derivative. Furthermore, the summation convention has been assumed.  

The stress tensor consists of four contributions, $\sigma_{ij}=\sigma_{ij}^{s}+\sigma_{ij}^{a}+ \sigma_{ij}^{Er}+\sigma_{ij}^{act}$ where 
\begin{gather} \label{eq:sigma1}
\sigma_{ij}^{s} = - P\delta_{ij}+2\eta v_{i,j}-\beta_1H_{ij}, \qquad\sigma^{a}= H_{ik} Q_{kj}-Q_{ik}H_{kj},\\
\sigma_{ij}^{Er} = -\frac{\p W}{\p Q_{kl,i}}Q_{kl,j},\qquad \sigma^{act}=-\zeta\delta\mu Q_{ij}.
\label{eq:sigma2}
\end{gather}
where $P$ is the hydrostatic pressure, $\eta$ the ordinary viscosity mentioned before, $\beta_1$ the rotational viscosity and $W$ the free energy density. The simplest form of this free energy density includes the elastic nematic energy within the 
the one-constant approximation and the Landau-de Gennes potential:
\be\label{eq:W}
W = K |\nabla {\bf Q}|^2+a \Tr({\bf Q})^2 +c \Tr({\bf Q})^4,
\ee
where $K$ and $c$ are positive constants, whereas $a$ changes sign at the temperature of the nematic--isotropic transition. 
The molecular tensor field ${\bf H}$ is the opposite of the functional derivative of the free energy density $W$ with respect to the order parameter ${\bf Q}$
\be\label{eq:H}
H_{ij} = -\frac{\delta W}{\delta Q_{ij}} = \(\frac{\p W}{\p Q_{ij,k}}\)_{,k}-\frac{\p W}{\p Q_{ij}}.
\ee
In the following we neglect $\sigma_{ij}^a$ and $\sigma_{ij}^{Er}$, because active gels are usually supposed to be poorly
ordered, namely $|{\bf Q}|\ll 1$, and thus only the lowest order terms are taken into account in eq.~(\ref{eq:sigma}).

The tensorial equation (\ref{eq:Q}) can be splitted into two scalar equations
\begin{gather}\label{eq:q}
\bv\cdot \nabla q  = \frac 1{\beta_2} \( K \big({\rm \Delta} q -4q |\nabla {\bp}|^2\big) - aq -\frac c2q^3\) +2{\beta_1} \bp\cdot(\nabla \bv)\bp-\beta_1\div \bv -\lambda\delta\mu q, \\ 
\label{eq:p}
q\bp\times (\nabla \bp,\bv)=\bp \times \[\frac K{\beta_2} \big[q{\rm \Delta} \bp + 2 (\nabla p)(\nabla q)\big ] +\beta_1(\nabla \bv)\bp\].
\end{gather}
The presence of active motors, represented by $\lambda\delta\mu$, leads to the renormalisation of the term linear in $q$.

\subsection{Circular geometry}
Let us first consider these equations in the polar coordinate system $\{r,\phi\}$. The Laplacian of the scalar order parameter $q$ becomes
\be
{\rm \Delta}q=\frac 1 r \p_r(r\p_r q)+\frac 1{r^2} \p_{\phi\phi} q.
\ee
We parametrise the director as $\bp=\cos\theta\bi_r+\sin\theta\bi_\phi$ and so we have 
\be
\nabla \bp = p_{rr}\bi_r\otimes\bi_r+p_{r\phi}\bi_r\otimes\bi_\phi+p_{\phi r}\bi_\phi\otimes\bi_r+ p_{\phi\phi}\bi_\phi\otimes\bi_\phi,
\ee
where
\begin{gather}
p_{rr}=\p_r(\cos\theta),\qquad p_{r\phi}=\frac 1r (\p_\phi\cos\theta-\sin\theta),\\
p_{\phi r}=\p_r(\sin\theta),\qquad p_{\phi\phi}=\frac 1r (\p_\phi\sin\theta+\cos\theta),\\
|\nabla \bp|^2=p_{rr}^2+p_{r \phi}^2+p_{\phi r}^2+p_{\phi\phi}^2=(\p_r\theta)^2+\frac 1{r^2}(\p_\phi\theta+1)^2.
\end{gather}

We consider first a velocity field in the general form $\bv=v_r \bi_r+v_\phi\bi_\phi$, yielding the velocity gradient tensor as
\be\label{eq:gradv}
\nabla \bv =\p_r v_r\, \bi_r\otimes\bi_r+ \frac 12\(\p_r v_\phi+ \frac1r(\p_\phi v_r - v_\phi) \)\, (\bi_r\otimes\bi_\phi+\bi_\phi\otimes\bi_r)+ \frac 1r (\p_\phi v_\phi + v_r)\,\bi_\phi\otimes\bi_\phi,
\ee
and thus
\begin{align}
\div \bv &=   \p_r v_r + \frac 1r (\p_\phi v_\phi + v_r),\\
\label{eq:pdvp}
\bp\cdot(\nabla \bv)\bp  & = \cos^2\theta\p_r v_r + \sin\theta\cos\theta \(\p_r v_\phi+\frac1r( \p_\phi v_r - v_\phi)\) + \frac {\sin^2\theta}r (\p_\phi v_\phi + v_r),\\ \label{eq:ptvp}
\bp\times(\nabla \bv)\bp & = \frac{\bi_z}2 \[ \cos2\theta\(\p_r v_\phi + \frac1r (\p_\phi v_r - v_\phi)\) + \sin2\theta\(-\p_r v_r+ \frac1r(\p_\phi v_\phi + v_r)\)\]. 
\end{align}
Then eqs.~(\ref{eq:q}) and (\ref{eq:p}) take the form
\begin{multline}\label{eq:qgen}
v_r\p_r q +\frac 1r v_\phi \p_\phi q =\frac 1{\beta_2} \[ K{\rm \Delta} q -4qK \(\big(\p_r\theta\big)^2 +\frac 1{r^2}\big(\p_\phi\theta+1\big)^2\)- q(a + \lambda\delta\mu\beta_2)-\frac c2 q^3\] +\\+ \beta_1 \[\cos 2\theta\( \p_r v_r -\frac 1r (\p_\phi v_\phi + v_r)\) + \sin2\theta \( \p_r v_\phi+ \frac1r( \p_\phi v_r - v_\phi)\)\],
\end{multline}
and 
\begin{multline}\label{eq:pgen}
q\(v_r \p_r\theta+ \frac {v_\phi}r (\p_\phi \theta+1)\)= \frac{1}{\beta_2}\[ Kq {\rm \Delta} \theta +2K\(\p_r \theta\p_r q +\frac 1{r^2}\p_\phi q (\p_\phi \theta+1)\)  \] +\\+ 
\frac{\beta_1}2 \[ \cos2\theta\(\p_r v_\phi + \frac1r (\p_\phi v_r - v_\phi)\) + \sin2\theta\(-\p_r v_r+ \frac1r(\p_\phi v_\phi + v_r)\)\].
\end{multline}

\begin{figure}
\centering
\includegraphics[width=0.8\linewidth]{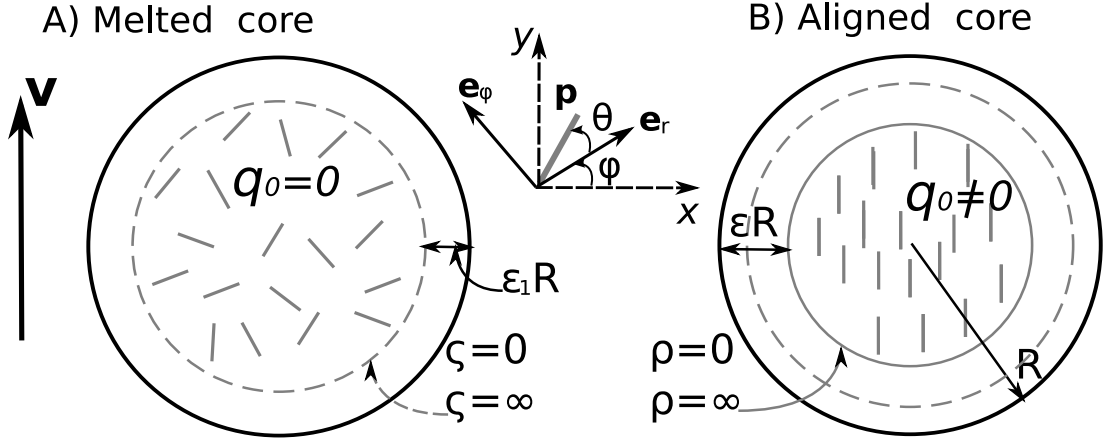}
\caption{Depending on the activity of motors $\lambda\delta\mu$ we can have two cases with A) melted core ($a+\lambda\delta\mu\beta_2>0$) or B) aligned core ($a+\lambda\delta\mu\beta_2<0$). In the case A) we define the width of a boundary layer ($\epsilon_1 R$) for the change of the degree of orientation $q$; in the case B) we define the boundary layer ($\epsilon R$) for the change of the director $\bp$.}
\label{fig:2cores}
\end{figure}

Both equations are essentially non-linear and the orientation of actin filaments is coupled to the flow, which will be found at the last section~\ref{sect:flow}. With the flow equation, we should deal  with three coupled non-linear  partial differential equations (P.D.E) for $q, \theta$ and $\Psi$, the stream function. There is no hope that these equations can be solved without assuming a certain relationships between the coefficients as well as the form of flow. At the boundary, the cell membrane together with immersed proteins imposes a preferred orientation on the actin filaments. Therefore, we will assign at the boundary a fixed angle $\theta_0$ between the normal and the fiber director $\bp$ and a fixed degree of orientation $\bar q$, which may depend locally on the shape of the membrane via its curvature. In a narrow region near the boundary we suppose that $q$ and $\theta$ change rapidly. Then the solution can be written in power series of $\epsilon$, which is a small dimensionless parameter, relating the width of a boundary layer, defined below, to the typical size of the cell $R$. To formulate the boundary layer problem, one needs to take into account non-linearities, and one expects to find the soliton-like behaviour of $q$ and $\theta$. Far away from the boundary, at the core of the cell, we have two cases: either isotropic state with `melted core' ($q_0=0$) or the uniformly aligned core with degree of orientation given by 
\be\label{eq:defq0}
q_0^2 = -\frac {c}{2(a+\lambda\delta\mu\beta_2)}>0.
\ee
These two cases are shown schematically in the fig.~\ref{fig:2cores}. Depending on the activity of motors $\lambda\delta\mu$ (positive or negative) the sign of $a+\lambda\delta\mu\beta_2$ can change. We will treat separately the cases with a melted core and an aligned core.

\section{The bulk and the actin cortex}

\subsection{Melted core: boundary layer for $q$}\label{subsec:melted}

Let us first consider a simple case, assuming that the angle $\theta$ is constant in the small region near the border, and its value is imposed by the cell membrane, so that $\theta=\theta_0$. Given that actin filaments approach the lipid membrane almost symmetrically, with some preferred angle either $\simeq20^\circ$ or $\simeq-20^\circ$~\cite{fletcher,andrew}, the average orientation at the border is along the normal, thus $\theta_0\simeq0^\circ$ and the nematic tensor ${\bf Q}$ is 
\be\label{eq:Qmelted}
Q_{rr}=-Q_{\phi\phi}=\frac q2, \quad \mbox{and} \quad  Q_{r\phi}=Q_{\phi r}=0.
\ee
Furthermore, we assume that the velocity field $\bv=v_r\,\bi_r + v_\phi\bi_\phi$ is small and can be disregarded for a moment (below we define a small parameter $\epsilon_1$ and the order of smallness for $v_r$ and $v_\phi$). Thus, eq.~(\ref{eq:qgen}) for the degree of orientation takes the form
\be\label{eq:qmelt}
K{\rm \Delta} q - q(a + \lambda\delta\mu\beta_2)-\frac c2 q^3=0.
\ee
Based on this equation we can introduce the dimensionless parameter 
\be\label{eq:epsilon1}
\epsilon_1=\frac 1R \sqrt{\frac K {|a+\lambda\delta\mu\beta_2|}},
\ee
relating the thickness of the boundary layer to the typical size of the cell $R$. In the case of  liquid crystals,  $\epsilon_1$ is the ratio of the nematic coherence length $\sqrt{K/a}$, which is of the order of nanometers, to the characteristic length  of the system $R\simeq1-10~\mu$m, yielding $\epsilon_1\lesssim 10^{-3}$. We can now estimate the strength of the advection term $v_r \partial_r q$ or $v_r \partial_r \theta$. Assuming that the main flow is given by $-k_d R/2$, the dimensionless parameter  $\beta_2k_d/a$ is of order   $\epsilon_1^2$.  In a thin region near the boundary, we suppose that $q$ can be written in power series of a small parameter $\epsilon_1$
\be\label{eq:qexp}
q = q^{(0)} + \epsilon_1 q^{(1)}\ldots
\ee
Let us introduce the rescaled variable $\varsigma$, 
\be
\frac rR =1 - \epsilon_1\varsigma,
\ee
measuring the distance from the boundary.  So $\varsigma=0$ corresponds to the border of the cell $r=R$, and $\varsigma \to \infty$ corresponds to the interior border of the boundary layer (see fig.~\ref{fig:2cores}A). At the leading order eq.~(\ref{eq:qmelt})  takes the form
\be\label{qq1}
\p_{\varsigma\varsigma}q^{(0)} - q^{(0)} - \omega (q^{(0)})^3 =0,\quad\mbox{where}\quad \omega=- \frac {c}{2(a+\lambda\delta\mu\beta_2)}>0,
\ee 
which should satisfy the asymptotic conditions 
\be\label{eq:bq0}
q^{(0)}|_{\varsigma\to \infty}=\p_{\varsigma} q^{(0)}|_{\varsigma\to \infty}=0,
\ee
and the boundary condition at the border of the cell
\be\label{eq:bq1}
q^{(0)}|_{\rho\to 0}=\bar q.
\ee
Multiplying both sides of eq.~(\ref{qq1}) by $\p_\varsigma q^{(0)}$ and integrating we find
\be
(\p_\varsigma q^{(0)})^2 - (q^{(0)})^2 - \frac{\omega}{2} (q^{(0)})^4 =0,
\ee
where the integration constant was chosen so to satisfy the asymptotic conditions (\ref{eq:bq0}). The solution of this equation admits the form 
\be\label{eq:order0}
q^{(0)}(\varsigma) =  \frac{\sqrt{2/\omega}}{\sinh (\varsigma + C)},  \quad C =  {\rm asinh}\(\frac{\sqrt {2/\omega}}{\bar q}\).
\ee
As expected, in the leading order we found a soliton-like behaviour for $q$. In the case of a non-circular cell, a local boundary layer analysis must be done with a coordinate frame defined from the normal and the tangent at the point of the interface $\partial \Omega$ considered. Then $\varsigma$ is simply the distance taken along the normal modified by the local curvature.  The set of coordinates is then $\varsigma$ and the arclength $s$. It turns out that for the leading order the equations are quite the same as the one derived from the cylindrical coordinate frame with some obvious adaptation which will be given without demonstration \cite{benamar}. In a local curvilinear coordinates the Laplacian up to the next order is given by $\p_{\varsigma\varsigma}-\kappa\p_{\varsigma}$, where $\kappa$ is the local curvature, $\kappa\equiv\p\vartheta/\p s\equiv-\p \phi/\p s$, chosen to be negative for convex interfaces. The components of velocity field $v_r,v_\phi$ being of the order of $\epsilon_1^2$, the next order term in the expansion eq.~(\ref{eq:qexp}) should satisfy the following equation  
\be
\label{order1}
\p_{\varsigma\varsigma}q^{(1)} - q^{(1)}\big(1+3\omega (q^{(0)})^2\big) = -\kappa\(1 +\frac{k_d R\beta_2}{2K} \) \p_\varsigma q^{(0)} .
\ee
For convergence of the series,  expansion represented by eq.~(\ref{eq:qexp}), we look for a solution of eq.~(\ref{order1}) with  vanishing $q^{(1)}$  for $\varsigma=0$ and $\varsigma\to\infty$. The adjoint theorem gives us the value of $\p_\varsigma q^{(1)}$  at  $\varsigma=0$  for such a solution, an information which is useful for future development of the analysis.

\subsection{Aligned core: boundary layer for $\bp$}
Now, the core of the actin network is ordered (see fig.~\ref{fig:2cores}B).  The coupling of the director's orientation with the flow, given by the term proportional to $\beta_1$ in eqs.~(\ref{eq:qgen}), (\ref{eq:pgen}), leads to the alignment of actin filaments in the direction of the flow. Without loss of generality we assume the flow to be in $y$-direction, as in section~\ref{sect:problem}. Therefore, the core of the cell is expected to be aligned in the same direction as shown in  fig.~\ref{fig:2cores}. The main velocity field is of the form 
$\bv = (V_0-k_d y)\bi_y$ with $\div \bv=-k_d$ (see section~\ref{sect:problem} and eq.~(\ref{div})). 
Note that at this stage, we do not have information on the value of  $V_0$. Then eqs.~(\ref{eq:qgen}) and (\ref{eq:pgen}) take the form
\begin{multline}\label{eq:q1}
(V_0-k_d r\sin\phi)\bigg(\sin\phi\p_r q +\frac 1r \cos\phi \p_\phi q \bigg)=\\= \frac 1{\beta_2} \[ K{\rm \Delta} q -4qK \(\big(\p_r\theta\big)^2 +\frac 1{r^2}\big(\p_\phi\theta+1\big)^2\)- q(a + \lambda\delta\mu\beta_2)-\frac c2 q^3\] +\beta_1k_d \cos 2 (\phi+\theta),
\end{multline}
and 
\begin{multline}\label{eq:p1}
q(V_0-k_d r\sin\phi)\bigg(\sin\phi\p_r\theta+\frac 1r \cos\phi (\p_\phi \theta+1)\bigg)=\\= \frac{1}{\beta_2}\[ Kq {\rm \Delta} \theta +2K\(\p_r \theta\p_r q +\frac 1{r^2}\p_\phi q (\p_\phi \theta+1)\)  \] -\beta_1\frac{k_d}2 \sin2(\phi+\theta).
\end{multline}

The boundary layer thickness ($\epsilon_1$) for the degree of orientation  was defined in eq.~(\ref{eq:epsilon1}), and it was a function of the activity of molecular motors $\lambda\delta\mu$. We may have different situations with a  double boundary layer structure, depending on the values of the coefficients in eqs.~(\ref{eq:q1}) and (\ref{eq:p1}). However, it is reasonable to assume that near the boundary the scalar order parameter $q$ changes more rapidly than the director. Thus we assume in this section that $q$ is constant, given by the equilibrium value in the bulk $q= q_0$ (eq.~(\ref{eq:defq0})) and $q_0/\beta_1$ is of the order of $\epsilon^2$, the dimensionless thickness of the boundary layer for the angle $\theta$,  defined as follows
\be\label{eq:epsilon}
\epsilon=\frac 1R\sqrt{\frac{q_0 K} {|\beta_1\beta_2| k_d}}.
\ee
The transport coefficients $\beta_1$ and $\beta_2$ can be written in terms of the rotational viscosity $\gamma_1>0$ and the torsion coefficient $\gamma_2$ as $\beta_1= -q_0\gamma_2/\gamma_1$, $\beta_2=\gamma_1/(2q_0^2)$~\cite{olmsted,degennes:book}. The torsion coefficient $\gamma_2=\eta_1-\eta_2$~\cite{stewart:book}, where $\eta_1$ is the Miesowicz viscosity when $\bp$ is parallel to $\bv$ and $\eta_2$ is  the Miesowicz viscosity when $\bp$ is parallel to $\nabla \bv$, can be positive (e.g. HBAB in nematic phase~\cite{pieranski}) or negative (e.g. MBBA in nematic phase~\cite{stewart:book}). Since we do not know neither the sign nor the value of $\gamma_2$ for actin filaments in cells, we used the absolute value of $\beta_1\beta_2=-\gamma_2/(2q_0)$ in eq.~(\ref{eq:epsilon}) and in the following assume $\gamma_2>0$. Taking into account that  $\epsilon_1<\epsilon$ (eqs.~(\ref{eq:epsilon1}), (\ref{eq:epsilon})), the following relationship between the coefficients  holds
\be\label{eq:small}
\epsilon_1<\epsilon \quad\Longleftrightarrow \quad |a+\lambda\delta\mu\beta_2|>|\beta_1\beta_2| k_d/q_0.
\ee
To estimate the value of $\epsilon$ we take the  elastic constant for liquid crystals $K\simeq 6 \cdot 10^{-12}$~N and assume that $|\beta_1\beta_2|\simeq \gamma_2\simeq\eta$, which is a poorly measured value, varying between $\eta\simeq 0.1$~Pa$\cdot$s~\cite{visco:pollard}, $\eta\simeq300$~Pa$\cdot$s~\cite{visco:science} and $\eta\simeq10^3-10^5$~Pa$\cdot$s~\cite{salbreaux:these}. Assuming the $\eta\simeq 10^4$~Pa$\cdot$s and $k_d\simeq 0.2$~1/s~\cite{andrew} we find  $\epsilon\simeq 0.01$ for the cell with radius $R=10~\mu$m and $\epsilon\simeq 0.1$ for $R=1~\mu$m. The smallness of $\epsilon$ justifies the assumed boundary layer approximation, however, the lack of information about the value of the torsion coefficient $\gamma_2$ and the rotational viscosity $\gamma_1$ remains a sensitive issue for our model.

We aim at solving the boundary layer problem for the director, by expanding the angle $\theta$ as a perturbative series
\begin{equation}\label{eq:thetaexp}
\theta=\theta^{(0)}+\epsilon\theta^{(1)}+\epsilon^2\theta^{(2)}+\ldots
\end{equation} 
Let us introduce the rescaled variable $\rho$
\be\label{eq:rho}
\frac rR = 1 -\epsilon\rho,
\ee
which measures the distance from the boundary, so that $\rho=0$ corresponds to $r=R$ and $\rho\to\infty$ corresponds to the interior border of the boundary layer (see fig.~\ref{fig:2cores}B).  We should satisfy the boundary conditions for the orientation at the zero-order approximation, so that
\be\label{eq:bc0}
\theta^{(0)}|_{\rho\to \infty}=\frac \pi2-\phi,\qquad \theta^{(0)}|_{\rho\to 0}=\theta_0,
\ee
whereas for the higher orders 
\be\label{eq:bc1}
\theta^{(i)}|_{\rho\to \infty}=\theta^{(i)}|_{\rho\to 0}=0,\qquad i=2,3,\ldots
\ee
The leading order approximation for the orientation satisfies the following equation
\begin{equation}\label{eq:theta0}
2\p_{\rho\rho} \theta^{(0)}+  \sin 2(\phi +\theta^{(0)}) = 0,
\end{equation}
with the first integral given by
\begin{equation}\label{eq:1int}
2 \p_{\rho\rho}{\theta^{(0)}} +  \sin 2(\phi+\theta^{(0)}) =0, \quad \to \quad (\p_{\rho}{\theta^{(0)}})^2 - \frac{\cos 2(\phi+\theta^{(0)})}2=C(\phi).
\end{equation}
The constant of integration $C(\phi)$ can be found from the  condition in the bulk eq.~(\ref{eq:bc0}), yielding $C(\phi)=1/2$ and consequently
\begin{equation}\label{eq:dtheta0}
(\p_{\rho}{\theta^{(0)}})^2  = \cos^2 (\phi+\theta^{(0)}).
\end{equation}
The further integration gives
\begin{equation}\label{eq:result0}
\theta^{(0)}= -\phi + 2\arctan\bigg(\frac{A(\phi) e^{\rho }-1}{A(\phi) e^{\rho}+1}\bigg), \qquad
A(\phi)= \frac{1+\tan\frac{\theta_0+\phi}2}{1-\tan\frac{\theta_0+\phi}2},
\end{equation}
where the constant $A(\phi)$ is chosen to satisfy the boundary condition at the membrane $\theta^{(0)}_{\rho\to0} = \theta_0$ (eq.~(\ref{eq:bc0})).

\begin{figure}
\centering
\includegraphics[width=0.47\linewidth]{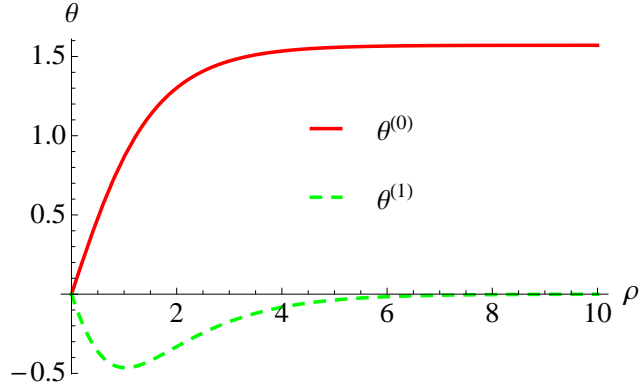}
\caption{The plot of $\theta^{(0)}$ and $\theta^{(1)}$ given by eqs.~(\ref{eq:result0}) and  (\ref{eq:result1}) as function of $\rho$ for $\phi=\theta_0=0$ and $\kappa=-1$.}
\label{fig:2func}
\end{figure}

To illustrate the convergence of the series (\ref{eq:thetaexp}) we need to find the next order term  in the expansion. By substituting eq.~(\ref{eq:thetaexp}) into  eq.~(\ref{eq:p1}), taking into account the contribution from the LHS, and collecting the terms proportional to $\epsilon$, the equation for $\theta^{(1)}$ takes the form
\begin{equation}\label{eq:theta1}
\p_{\rho\rho}\theta^{(1)} +\theta^{(1)}\cos2(\phi+\theta^{(0)})=-\kappa\p_\rho \theta^{(0)} \underbrace{\(1-\frac{q_0\sin^2\phi}{|\beta_1|}+\frac{q_0V_0\sin\phi}{k_d|\beta_1|R}\)}_{={\cal V}(\phi)}.
\end{equation}
Substituting the results for $\theta^{(0)}$, eq.~(\ref{eq:result0}), in eq.~(\ref{eq:theta1}) we find
the equation for $\theta^{(1)}$
\begin{equation}
\p_{\rho\rho} \theta^{(1)} + \theta^{(1)} \bigg[2 \bigg(\frac{2A e^{\rho}}{1+A^2 e^{2\rho}}\bigg)^2 -1\bigg] = -\kappa\frac{A e^{\rho}}{1+A^2 e^{2\rho}}{\cal V}(\phi),
\end{equation}
and the solution in a compact form
\begin{equation}\label{eq:result1}
\theta^{(1)}=\kappa {\cal V}(\phi)\, \frac{A^2 e^{\rho} \rho + \sinh \rho}{2A(1 + A^2 e^{2 \rho})}.
\end{equation}
which satisfies the boundary conditions $\theta^{(1)}(0)=\theta^{(1)}(\infty)=0$. In fig.~\ref{fig:2func} we compare the first two terms of the expansion of $\theta$ (\ref{eq:thetaexp}) and notice that the amplitude of  $\theta^{(1)}$ is smaller than $\theta^{(0)}$ in the whole range of $\rho$. Nevertheless, eq. (\ref{eq:result1}) is valid if and only if $q$ is kept constant up to order $\epsilon^2$ in this boundary layer. This restriction which is absolutely not necessary deprives us of a degree of freedom which would be very useful for the next orders of the asymptotics when the flow field and the shape of the cell is considered. A much better strategy will be to solve eq. (\ref{eq:q1}) and eq. (\ref{eq:p1}) by keeping the expansion eq. (\ref{eq:thetaexp}) for $\theta$ but the following expansion for $q$
\begin{equation}
q=q_0\big(1+\epsilon q^{(1)}\big).
\end{equation}
This viewpoint does not modify our leading order and  we consider now the modification of the hydrodynamic equations induced by  the actin cortex.

\section{Flow in the actin cortex}\label{sect:flow}

Actin cortex exerts forces on a lipid membrane, resulting in the modification of the boundary conditions (eqs.~(\ref{neu}) and (\ref{dir})) and consequently the shape of the drop, eq.~(\ref{schwarz}). Let us consider the balance of forces given by eq.~(\ref{eq:sigma}). Since we are in the limit of the boundary layer we adopt a Cartesian system of coordinates. Later on, we will change to a local coordinate system more suitable to fix the boundary conditions. Taking the divergence of this tensorial equation, we find a modified Stokes equation (see eq.~(\ref{stokes})), 
\be\label{eq:vi}
\xi v_i + \frac{\p P} {\p x_i} = \eta{\rm \Delta} v_i +\frac{\p\Lambda_{ij}}{\p x_j}, 
\ee  
where 
\be\label{eq:lambda}
\Lambda_{ij}=-\zeta\delta\mu Q_{ij}-2\beta_1 (KQ_{ij,kk}-aQ_{ij}-2cQ_{ij}Q_{kl}Q_{kl}),
\ee
is the stress tensor (see eqs.~(\ref{eq:sigma1}), (\ref{eq:sigma2}) and (\ref{eq:H})), accounting only for the lowest order contributions as was referred above. Introducing the stream function $\Psi$  as before in eq.~(\ref{eq:stream}), so that 
\be\label{eq:vxvy}
v_x =\frac{\p \Psi}{\p y},\qquad v_y=V_0-k_dy - \frac{\p \Psi}{\p x}, 
\ee
we can rewrite the eq.~(\ref{eq:vi}) in the form
\begin{align}\label{eq:vx}
\xi \frac{\p \Psi}{\p y} + \frac{\p P} {\p x} &= \eta{\rm \Delta}\frac{\p \Psi}{\p y}+ \nabla {\rm \Lambda}|_{x},\\
-\xi \(k_d y  + V_0+ \frac{\p \Psi}{\p x}\) + \frac{\p P} {\p y} &= -\eta{\rm \Delta}\frac{\p \Psi}{\p x}+ \nabla {\rm \Lambda}|_{y},\label{eq:vy}
\end{align}
where $\nabla {\rm \Lambda}|_{i} =\Lambda_{ij,j}$. Taking the cross derivatives of the above equations ($\p_y {\rm eq.}~(\ref{eq:vx})-\p_x \rm{eq.}~(\ref{eq:vy})$) we eliminate the pressure $P$ and find 
\be\label{eq:Psivec}
\xi {\rm \Delta} \Psi = \eta {\rm \Delta}^2 \Psi + \p_y \nabla {\rm \Lambda}|_{x} - \p_x \nabla {\rm \Lambda}|_{y}.
\ee
This equation will replace the hydrodynamic equation~(\ref{eq:sigma}). One can also recognize the part of eq.~(\ref{complete}) derived in the section~\ref{sect:problem}.  As previously,  we neglect the viscosity as was already stated in section~\ref{sect:problem}, so $\eta=0$. Note that in this limit far away from the actin cortex  ${\bf \Lambda}|_{\rho\to\infty}\sim0$, one recovers the Laplace equation for the stream function and therefore the velocity potential formulation considered in the section~\ref{sect:problem}.

Coming back to the polar system of coordinates, so that $\{x,y\}\to\{r,\phi\}$, we have
\begin{align}
\nabla {\rm \Lambda}|_{r} &= \frac{\p \Lambda_{rr}}{\p r} +\frac 1r \(\frac{\p \Lambda_{r\phi}}{\p \phi}+\Lambda_{rr}-\Lambda_{\phi\phi}\),\label{eq:lambda_r}\\
\nabla {\rm \Lambda}|_{\phi} &= \frac{\p \Lambda_{\phi r}}{\p r} +\frac 1r \(\frac{\p \Lambda_{\phi\phi}}{\p \phi}+\Lambda_{\phi r} + \Lambda_{r \phi}\)\label{eq:lambda_phi},
\end{align}
yielding eq.~(\ref{eq:Psivec}) in the case $\eta=0$ as
\be\label{eq:Psi}
\xi {\rm \Delta} \Psi = \frac 1r \frac{\p^2\Lambda_{rr}}{\p r \p\phi} + \frac 1{r^2} \(\frac{\p^2\Lambda_{r\phi}}{\p \phi^2} +\frac{\p\Lambda_{rr}}{\p \phi }-\frac{\p\Lambda_{\phi\phi}}{\p \phi }\) - \frac 1 r \frac \p {\p r} \( r \frac{\p\Lambda_{\phi r}}{\p r}\) - \frac 1r\frac{\p^2\Lambda_{\phi\phi}}{\p r\p \phi}-\frac 1 r \frac{\p\Lambda_{\phi r}}{\p r}-\frac 1 r \frac{\p\Lambda_{r \phi}}{\p r}.
\ee

Substituting the equilibrium value of the scalar order parameter $q_0$ (eq.~(\ref{eq:defq0})) into eq.~({\ref{eq:lambda}}) we get
\be\label{eq:lambda1}
\Lambda_{ij} = -(\zeta\delta\mu +2\beta_1\beta_2\lambda\delta\mu)Q_{ij} -2\beta_1KQ_{ij,kk}.
\ee
This implies that the Laplace law of capillarity is modified by adding $\Lambda_{rr}$. Next, based on eq.~(\ref{eq:small}) we compare two terms in eq.~(\ref{eq:lambda1}), and find that when $\lambda\delta\mu\gg k_d$  we can account only for the active stress and omit the last term which is elastic. 
Indeed, according to experimental observations~\cite{yam}, the motility of cell can be related to the increase of the activity of the motors, therefore it is reasonable to assume that $\lambda \delta\mu$ increases, yielding only the term linear in $Q_{ij}$ within our boundary layer  approximation
\be\label{eq:approx}
\Lambda_{ij} \simeq -Q_{ij}(\zeta\delta\mu +2\beta_1\beta_2\lambda\delta\mu).
\ee

\begin{figure}
\centering
A) $k_d=V_0=0$\hfil B) $k_d\neq 0$\hfil C) $V_0\neq 0$\\
\includegraphics[height=50mm]{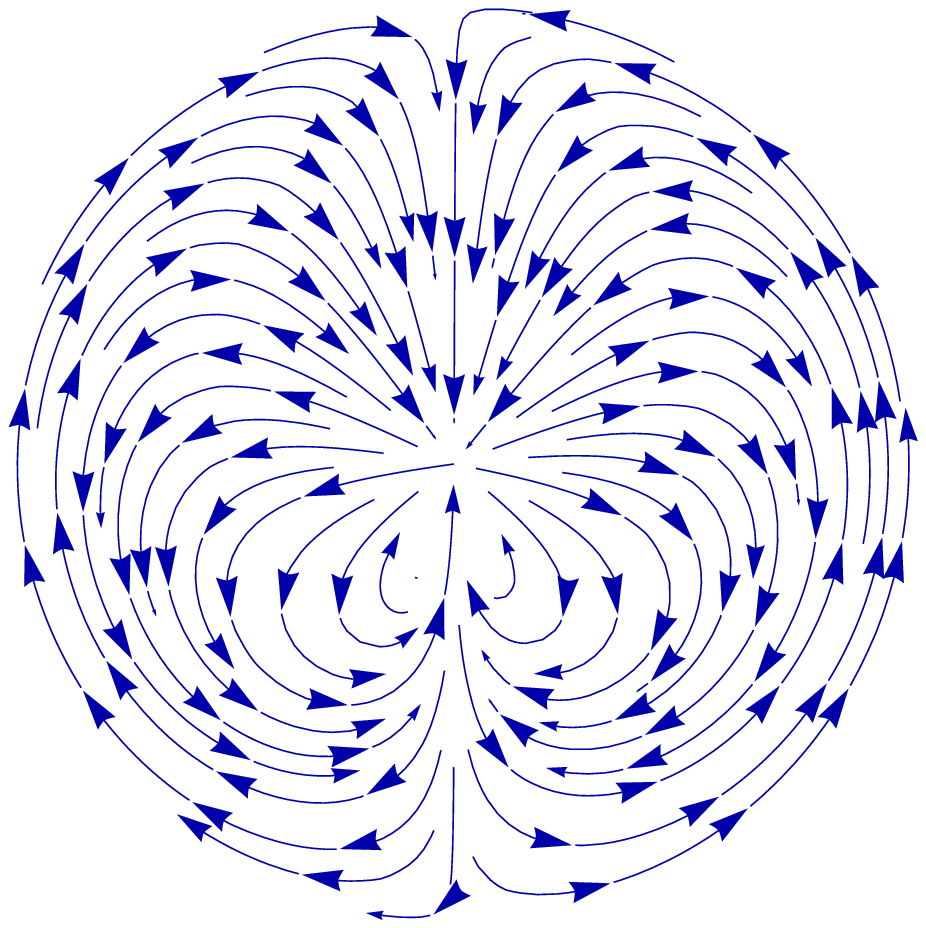}\hfil
\kern-10pt\includegraphics[height=50mm]{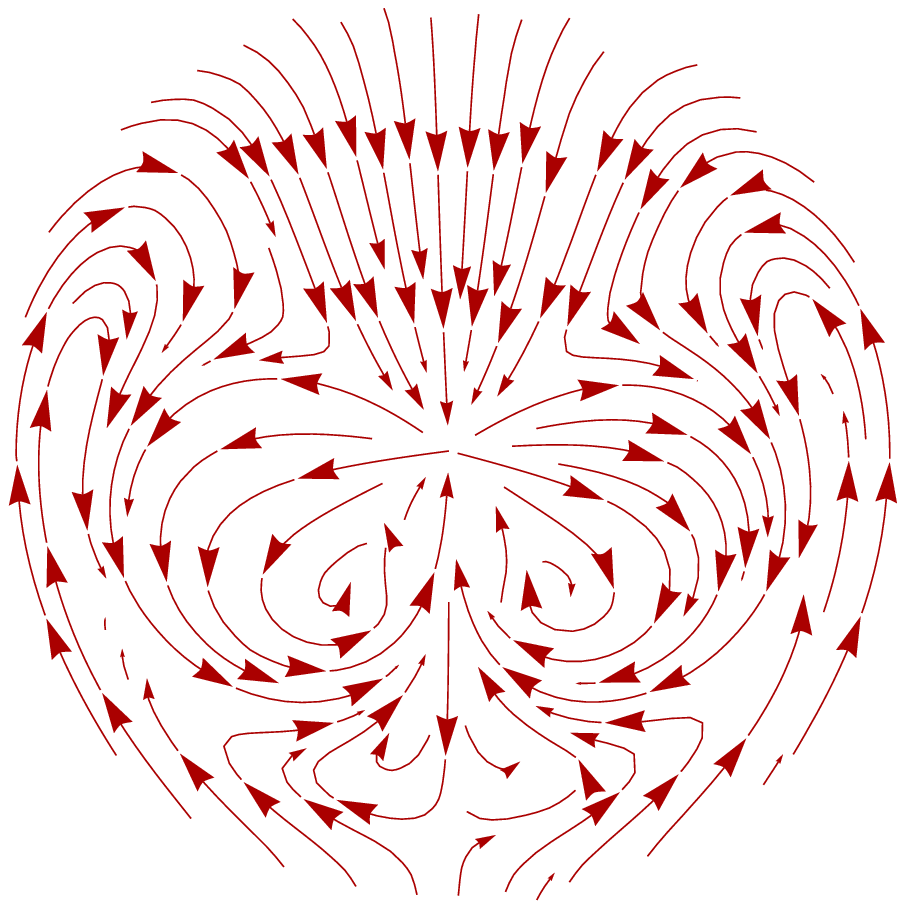}\hfil
\kern-10pt\includegraphics[height=50mm]{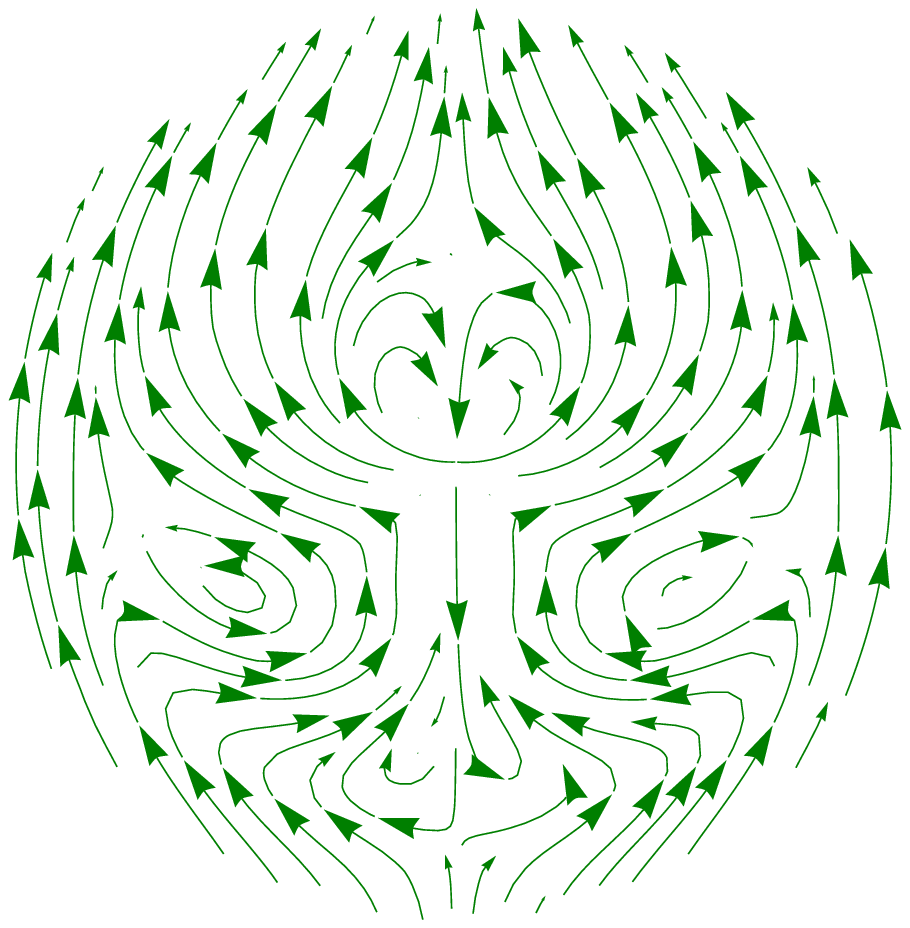}
\caption{Flow in the actin cortex for $(v_x,v_y)$, given by eq.~(\ref{eq:vxvy}) for the stream function $\Psi$ (eq.~\ref{eq:Psi1}) only (A) and for $\Psi$  supplemented with a non-zero value of $k_d$ (B) or a non-zero value of $V_0$ (C), which are the dominant contributions. The meaningful part of the flow is only near the border, where the actin cortex is localised.}
\label{fig:flow}
\end{figure}

We will solve eq.~(\ref{eq:Psi}) for the stream function  within the boundary layer approximation. Since in section~\ref{sect:nematic} we assumed the leading order contribution to the flow in the form $\bv=(V_0-k_d y)\,\bi_y$, the stream function $\Psi$, representing here the higher order corrections, takes the form
\be
\Psi=\epsilon^3 \Psi^{(1)} + \epsilon^4\Psi^{(2)}+\ldots
\ee 
Substituting this expansion together with eq.~(\ref{eq:approx}) into eq.~(\ref{eq:Psi}) and introducing the parameter $\tilde\xi = \xi/(\zeta\delta\mu +2\beta_1\beta_2\lambda\delta\mu)$, so that $\xi\sim O(1/\epsilon^3)$, which is indeed the limit of  strong friction assumed above, we find 
\be
\tilde\xi \big (\p_{\rho\rho}\Psi^{(1)}+\epsilon \p_{\rho\rho} \Psi^{(2)} + \epsilon\kappa \p_{\rho}\Psi^{(1)} \big) = \p_{\rho\rho}Q_{r\phi}^{(0)} + \epsilon \p_{\rho\rho}Q_{r\phi}^{(1)} + 3\epsilon \kappa\p_{\rho}Q_{r\phi}^{(0)} +2\epsilon\p_{\rho s}Q_{rr}^{(0)},
\ee 
where we used the fact that  ${\bf Q}$ is a symmetric and traceless tensor, and we have introduced a local curvilinear system of coordinates as above with the local curvature $\kappa\equiv\p\vartheta/\p s\equiv-\p \phi/\p s$, negative for convex interfaces. The leading order contribution to the stream function is
\be\label{eq:Psi1}
\Psi^{(1)} = \frac{Q_{r\phi}^{(0)}}{\tilde \xi} = \frac{q_0} {2\tilde \xi}\sin(2\theta^{(0)}),
\ee
where $q_0$ and $\theta^{(0)}$  are given by eqs.~(\ref{eq:defq0}) and (\ref{eq:result0}), respectively. Taking into account this correction to the main velocity field $\bv = (V_0-k_d y)\,\bi_y$, considered in 
the section~\ref{sect:nematic}, in fig.~\ref{fig:flow} we visualise the flow given by eq.~(\ref{eq:vxvy}). 
The single contribution of the stream function $\Psi$ ((eq.~\ref{eq:Psi1}), fig.~\ref{fig:flow}A) suggests that the flow near the border, where the actin cortex is localised, has a preferred $y$-direction. Therefore, we may expect to find the motility of the cell as will be shown in the next section. In the case of the melted core $Q_{r\phi}=0$ (eq.~(\ref{eq:Qmelted})) and we find 
\be\label{eq:Psi2}
\frac{\p \Psi^{(2)}}{\p \rho} =\frac{2}{\tilde \xi} \frac{\p Q_{rr}}{\p s}.
\ee

\section{Boundary conditions}\label{sect:bound}

Due to the Cauchy relations
\be
\bn\cdot \nabla\Phi= -{\bf t}\cdot \nabla\Psi, \qquad {\bf t}\cdot \nabla\Phi= {\bf n}\cdot \nabla\Psi,
\ee 
we can obtain the velocity potential from the stream function
\be
\Phi=\Phi^{(0)}+\epsilon^2 \Phi^{(1)}+\epsilon^3 \Phi^{(2)}+\epsilon^4\Phi^{(3)}+\ldots,  
\ee
where $\Phi^{(0)}$ is given by eq.~(\ref{eq:Phi1}) used in the  section~\ref{sect:problem} to formulate the boundary conditions eq.~(\ref{eq:Phi0}) and being of order $\epsilon^2$. The next order contribution to the boundary conditions is calculated by taking into account eq.~(\ref{eq:Psi1}), resulting in
\begin{align}
\frac{\p\Phi^{(1)}}{\p r}&= -  \epsilon\frac{\p \Psi^{(1)}}{\p s} = \epsilon\frac {\vert \kappa\vert q_0}{2\tilde\xi}  \frac{\p\sin(2\theta^{(0)})}{\p\phi}\bigg|_{\rho\to 0} = 0,\\[1ex]
\frac{\p\Phi^{(1)}}{\p s}&=  \epsilon\frac{\p \Psi^{(1)}}{\p r} = -\frac {q_0} {2R\tilde\xi} 
\frac{\p\sin(2\theta^{(0)})}{\p\rho}\bigg|_{\rho\to 0} = -\frac {q_0} {R\tilde\xi} \cos(\theta_0 + \phi) \cos(2 \theta_0).
\end{align}

To sum up we have obtained a modification of the Neumann boundary condition: $\Phi^{(2)}=-\bar q/(R \tilde \xi)$ which has some consequence for the melted core solution if and only if
the prescribed value of $\bar q$ at the boundary depends explicitly on the arclength. In the absence of biological information, we have taken this value constant along the cell boundary and this contribution does not modify the free boundary problem. In the case of the oriented core we have a modification of the Neumann condition and we can evaluate its consequence on eq.~(\ref{schwarz}).
Cancelling  the value of $\theta_0$ (averaged value) and noticing that $\phi$ has to be identified to $\pi/2 -\vartheta$ the left-hand-side of eq.~(\ref{neu}) has the following additive contribution given by
\be
\frac{q_0} {R\tilde\xi}\sin(\vartheta)=\frac{q_0} {2iR\tilde\xi}\( \frac{1}{\sqrt{g'(z)}} - \sqrt{g'(z)} \). 
\ee
As a consequence the left-hand side of eq.~(\ref{schwarz}) is also modified by a term which is $-i q_0  (1-g'(z))/(R{\tilde\xi})$ which compensates exactly the singularity induced by the velocity and we derive:
\be\label{eq:U}
U = -\frac{q_0} {R\tilde\xi}=-\frac{q_0}{R\xi}\(\zeta\delta\mu +2\beta_1\beta_2\lambda\delta\mu\).
\ee
First we note that the velocity $U$  is positive for contractile motors in agreement with experiments. We can roughly  estimate the value of this translational displacement of the cell. Taking the averaged orientation of actin filaments at the border as zero, $\theta_0\simeq 0$, the radius of the cell $R\simeq 1-10~\mu$m, the value for the measured friction coefficient in keratocytes $\xi\simeq 10^5$~Pa$\cdot$s$/\mu^2$~\cite{andrew}, and the activity of motors $\zeta\delta\mu\simeq -10^3$~Pa, corresponding to contractile motors~\cite{andrew} we get an estimate of the first term in eq.~(\ref{eq:U}) as $U\sim 1-10$~nm/s. We also assumed a perfect ordering in the center of a cell, $q_0=1$ in eq.~(\ref{eq:defq0}). The second term can be rewritten taking into account that $2\beta_1\beta_2q_0=-\gamma_2$, which gives $\lambda\delta\mu\gamma_2/(R\xi)$. The viscosity of actin filaments $\eta\simeq 10^4$~Pa$\cdot$s~\cite{andrew,salbreaux:these}, which is orders of magnitude higher than for any nematic liquid crystal. Assuming $\gamma_2\simeq\eta$ and  $\lambda\delta\mu\simeq 5 k_d\simeq 1$~s$^{-1}$, we find the velocity $U\sim 10$~nm/s of the same order as above.

\section{Conclusions}

We have shown that it is possible to explain the cell motility through the reorganisation of the actin network in a two-dimensional cell in the limit of strong friction interaction with a substrate. The cytoskeleton dynamics is described by the hydrodynamics equations for active gels. Due to polymerisation occurring below the lipid membrane and depolymerisation at the other end of the filament, a flow of F-actin filaments exists  that we modelled by a Darcy flow. The fact that the organisation of the cortex imposed by the membrane is limited to a short distance from the border allows to treat the  equations of active gels for an arbitrary cell geometry.  Nevertheless, to leading order in the thickness boundary layer, we have shown that a circular cell with no order orientation for the F-actin filaments  in the bulk is static while an  averaged longitudinal order induces an hydrodynamic flow compatible with the model (bulk hydrodynamic and boundary conditions).  Modification of the cell shape can be obtained by extending the present work to next orders and by introducing biological informations necessary to improve the description.

\begin{acknowledgement}
It is a pleasure to thank fruitful and enlightful conversations with Andrew Callan-Jones, Pasquale Ciarletta, Linda Cummings and  Jean-Fran\c{c}ois Joanny. O. M. was supported by the French National Research Agency (ANR), grant ANR-07-BLAN-0158. G. N. was partly supported by University ``Pierre et Marie Curie''.
\end{acknowledgement}
%

\end{document}